# Modelling the Inner Debris Disc of HR 8799

Bruna Contro[1,2], Jonti Horner[3,4], Rob Wittenmyer[2,3,4], Jonathan P. Marshall[2,4], T. C. Hinse[5,6,3]

[1]*University of Sao Paulo State, Sao Paulo, Botucatu 18618-970, Brazil*
[2]*School of Physics, UNSW Australia, Sydney, New South Wales 2052, Australia*
[3]*Computational Engineering and Science Research Centre, University of Southern Queensland, Toowoomba, Queensland 4350, Australia*
[4]*Australian Centre for Astrobiology, UNSW Australia, Sydney, New South Wales 2052, Australia*
[5]*Korea Astronomy and Space Science Institute, 776 Daedukdae-ro, Yuseong-gu, Daejeon 305-348, Republic of Korea.*
[6]*Armagh Observatory, College Hill, Armagh BT61 9DG, UK*

**Abstract:**
In many ways, the HR 8799 planetary system strongly resembles our own. It features four giant planets and two debris belts, analogues to the Asteroid and Edgeworth-Kuiper belts. Here, we present the results of dynamical simulations of HR8799's inner debris belt, to study its structure and collisional environment.

Our results suggest that HR 8799's inner belt is highly structured, with gaps between regions of dynamical stability. The belt is likely constrained between sharp inner and outer edges, located at ~6 and ~8 au, respectively. Its inner edge coincides with a broad gap cleared by the 4:1 mean-motion resonance with HR 8799e.

Within the belt, planetesimals are undergoing a process of collisional attrition like that observed in the Asteroid belt. However, whilst the mean collision velocity in the Asteroid belt exceeds 5 kms$^{-1}$, the majority of collisions within HR 8799's inner belt occur with velocities of order 1.2 kms$^{-1}$, or less. Despite this, they remain sufficiently energetic to be destructive – giving a source for the warm dust detected in the system.

Interior to the inner belt, test particles remain dynamically unstirred, aside from narrow bands excited by distant high-order resonances with HR 8799e. This lack of stirring is consistent with earlier thermal modelling of HR 8799's infrared excess, which predicted little dust inside 6 au. The inner system is sufficiently stable and unstirred that the formation of telluric planets is feasible, although such planets would doubtless be subject to a punitive impact regime, given the intense collisional grinding required in the inner belt to generate the observed infrared excess.

## 1. Introduction

In the past two decades, a rapidly increasing number of planets have been detected orbiting other stars (3268 planets, according to the NASA Exoplanet Archive, on 20$^{th}$ May, 2016)[1]. The first to be found orbiting Sun-like stars were massive, giant planets, orbiting very close to their host stars (as exemplified by the first such planet discovered: 51 Pegasi b: Mayor & Queloz, 1995). Indeed, such planets remain the most easily detected, a direct result of the bias native to the two principal means of exoplanet detection: radial velocity and transit observations[2]. As time has passed, however, astronomers have gradually pushed the bounds of known exoplanets to ever smaller and more distant worlds (e.g. Jones et al., 2010; Delfosse et al., 2013; Wittenmyer et al., 2014a), with the Kepler spacecraft (Borucki et al., 2010) discovering over 1000 new planets in the last few years[3] (e.g. Borucki et al., 2011; Howard et al., 2012; Batalha et al., 2013; Rowe et al., 2014; Mullally et al., 2015).

---
[1] Statistics on the exact number of confirmed exoplanets are kept by three online databases. The precise number
[2] A detailed review of the various techniques used to detect exoplanets is beyond the scope of this work. We direct the interested reader to e.g. Perryman (2014), for more information.

In recent years, as the catalogue of known exoplanets has grown, the focus of planet search programs has shifted to the search for Solar system analogues. As radial velocity surveys have obtained longer temporal baselines, ever more distant planets have been found (e.g. Howard et al., 2010; Wittenmyer et al., 2012; Robertson et al., 2012). This process has culminated with the announcement of the first Jupiter analogues – giant planets moving on orbits similar to that of the giant planet Jupiter in our own Solar system (e.g. Boisse et al., 2012; Zechmeister et al., 2013; Wittenmyer et al., 2013, 2014b). These results have been complemented by those from Direct Imaging surveys, which have also begun to yield the discovery of massive planets at large astrocentric distances (e.g. Marois et al., 2008, 2010; Lagrange et al., 2009; Macintosh et al., 2015). The search for planetary systems that truly resemble the Solar system is important as it enables astronomers to better understand the uniqueness of our own planetary system. In addition, it is partially motivated by the search for habitable planets orbiting other stars (e.g. Horner & Jones, 2010, Stark et al., 2014).

For a system to be truly analogous to the Solar system, it needs to have more than just a Jupiter-analogue planet. The Solar system contains a plethora of small objects, in addition to the eight planets. The great bulk of these are stored in highly populous reservoirs – the Asteroid belt (e.g. Tedesco & Desert, 2002; Bottke et al., 2005), the Jovian and Neptunian Trojans (e.g. Jewitt, Trujillo & Luu, 2000; Sheppard & Trujillo, 2006), the trans-Neptunian disc (e.g. Jewitt & Luu, 1993; Brown, 2001; Gladman et al., 2001), and the Oort cloud (e.g. Dones et al., 2004). The structure of these reservoirs is highly complex, as is best illustrated by the case of the Asteroid belt.

The Asteroid belt stretches from an inner edge located just beyond the orbit of Mars to a sharp outer edge at around 3.25 au, well within the orbit of Jupiter. Within its broad torus, the belt contains asteroids on orbits that range from the circular to the moderately eccentric (several asteroids in the main belt have eccentricities approaching, or exceeding 0.5). At the same time, there are regions in the belt where objects are concentrated at moderate (~20 degrees) or high (~40 degrees, or more) orbital inclinations. Furthermore, the asteroids within the belt are neither smoothly nor continuously distributed between the inner and outer edges. When the semi-major axes of the main belt asteroids are plotted, rarefactions are readily seen. These are the Kirkwood gaps, located at semi-major axes where objects would have orbital periods in an integer ratio with that of Jupiter (a situation known as mean-motion resonance). In addition, a small population of asteroids lurk beyond the outer edge of the main belt, located at semi-major axes of around 4 au. These objects are known as the Hildas (e.g. Nesvorný & Ferrz-Mello, 1997), and are trapped in 3:2 mean-motion resonance with Jupiter, completing three orbits of the Sun for every two completed by the giant planet. The distribution of objects in the Asteroid belt can be seen in Figure 1.

---

[3] As of 20th May 2016, the tally of confirmed Kepler planets stands at 2327 (http://kepler.nasa.gov).

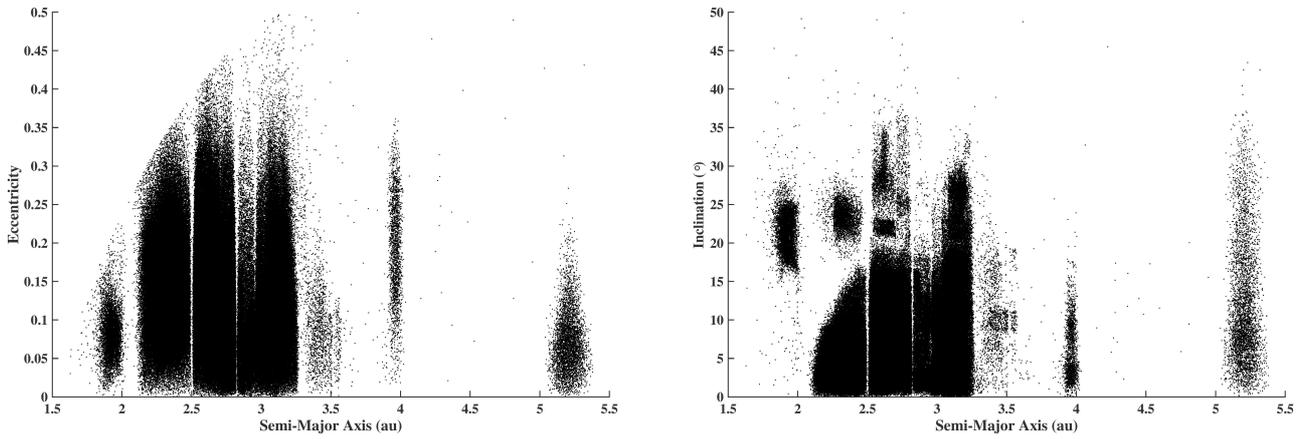

*Figure 1*: *The distribution of objects in the asteroid belt. The left panel shows the location of each numbered asteroid as a function of semi-major axis and eccentricity, whilst the right panel shows their locations as a function of semi-major axis and inclination. Data taken from the MPC Orbit Database, download on 13$^{th}$ October 2015 from http://www.minorplanetcenter.net/iau/MPCORB.html. The fine structure evident in both plots is the result of the dynamical influence of the giant planets.*

The fine structure of the Asteroid belt is the direct result of the gravitational influence of the giant planets. As mentioned above, the Kirkwood gaps are located at various resonant locations, places at which periodic perturbations by Jupiter accumulate to excite and eventually remove asteroids from the belt. Equally, other resonances (such as the 3:2 and 1:1 Jovian MMRs, occupied by the Hildas and Jovian Trojans, respectively) are dynamical 'sanctuaries' for objects, acting to protect objects from ejection, and therefore hosting large populations. The degree to which the Asteroid belt is excited (with large numbers of eccentric and inclined members), and sculpted tells the tale of the Solar systems' youth, when the belt was perturbed by the migration of the giant planets (e.g. Minton & Malhotra, 2009, Morbidelli et al., 2010).

The objects in the Asteroid belt, and other stable reservoirs of Solar system small bodies, have played an important role in determining the habitability of the Earth – playing a role in the delivery of our planet's volatiles (e.g. Owen & Bar-Nun, 1995; Morbidelli et al., 2000; Horner et al., 2009), and also contributing to the ongoing cycle of impact and extinction experienced by our planet (e.g. Alvarez et al., 1980; Becker et al., 2001; Horner & Jones, 2008, 2009). Over time, the Asteroid belt and trans-Neptunian disc are being ground to dust. Each collision produces vast quantities of dust, which is removed from the Solar system on short timescales by a variety of non-gravitational effects (such as Poynting-Robertson drag and radiation pressure; e.g. Wyatt & Whipple, 1950, Lagrange et al., 1995, Wyatt et al., 2011) – a process that could render our debris belts visible to observers around other stars, should they possess sufficiently sensitive infrared observatories.

We have known for over thirty years that our Solar system is not the only one to feature large populations of small bodies. The first 'debris discs' around other stars were detected by chance by the InfraRed Astronomical Satellite, IRAS (Neugebauer et al., 1984). That satellite detected excess infrared emission from the bright stars Vega, Fomalhaut, and Beta Pictoris – each of which was significantly more luminous at far-infrared wavelengths than was expected (Aumann et al., 1984; Aumann, 1985; Backman & Paresce, 1993). The only viable explanation for that excess was the presence of large quantities of dust in orbit around those stars, absorbing their light and re-radiating it at infrared wavelengths. Given that circumstellar dust is efficiently removed by radiation processes (e.g. Burns, Lamy & Soter, 1979; Bachman & Paresce, 1993; Krivov, 2010), we infer the presence of an unseen population of larger, asteroidal, bodies to replenish the visible dust through mutual collisions.

In the thirty years since IRAS, the number of known debris discs has, like the number of known planets, grown dramatically. Major contributions to expanding the ensemble of debris discs have been made by infrared satellite observatories such as ISO (e.g. Habing et al. 2001), Spitzer (e.g. Beichman et al. 2006; Su et al. 2006; Trilling et al. 2008) and WISE (Krivov et al. 2011, Kennedy & Wyatt 2013, Patel, Metchev & Heinze, 2014). Most recently, the Herschel Space Observatory (Pilbratt et al. 2010), with its large, 3.5-m primary mirror, has helped us to better characterise these discs through the measurement of resolved or extended emission for the first time (e.g. Liseau et al. 2010; Marshall et al. 2011; Löhne et al. 2012; Lestrade et al., 2012; Booth et al. 2013; Duchêne et al., 2014; Ertel et al. 2014; Faramaz et al. 2014; Pawellek et al. 2014; Marshall et al. 2014).

Assuming a system in thermal equilibrium, a dust belt that lies closer to its host star will be warmer than one farther away. Warmer dust exhibits measureable excess at shorter wavelengths (typically the mid-infrared), although the detection of such dust is limited at shorter wavelengths by contrast against the contribution of the stellar photosphere to the total emission. From the dust temperature, derived by fitting a blackbody curve to the measured excess, we can obtain an estimate of the dust location around its host star (e.g. Wyatt, 2008). This simple picture is complicated by the fact that smaller dust grains will typically be hotter than their larger brethren. In addition, those small grains move and disperse more rapidly – so the small grained component of a given belt's dust will be more widely spread, and hotter, than the larger-grained component. Despite this added complexity, it is now possible for us to begin to search for systems that are analogous to our Solar system in the distribution of the small body populations, as well as their planets – and understanding both these factors will prove key in determining the habitability of any Earth-like planets we find in those systems in the future (e.g. Horner & Jones, 2010, and references therein).

In the search for true Solar system analogues, then, the 'holy grail' is a system with a number of outer giant planets (analogous to Jupiter, Saturn, Uranus and Neptune), and debris belts that resemble those seen in our Solar system. The detection of Earth-like planets in the habitable zone currently lies beyond our reach, but near-future projects such as the Transiting Exoplanet Survey Satellite, TESS (Ricker et al., 2015), Plato (Rauer et al., 2014), and ground-based surveys such as MINERVA (Swift et al., 2015) should provide this capability for the nearest and brightest stars. Of the many known exoplanetary systems, the one that currently best fits this profile is that around the young A-star, HR 8799.

In 2008, the discovery of the first three giant planets orbiting HR 8799 was announced (Marois et al., 2008). Unlike the great majority of known exoplanets, those around HR 8799 were directly imaged using the Keck and Gemini telescopes (as well as a pre-discovery image obtained by Subaru, Fukagawa et al., 2009), allowing the authors to find planets far more distant than those that found through Radial Velocity and Transit surveys. The planets, only visible due to the youth of the system (still hot from their formation) were estimated to have masses between five and thirteen times that of Jupiter, and to lie approximately 24, 38 and 68 astronomical units from their host star, respectively. The discovery of a fourth planet (HR 8799 e; at ~15 au) was announced by the same group in 2010, found once again on the basis of direct images (Marois et al. 2010). Dynamical studies of the system have revealed that the planets probably move on near-circular orbits (e.g. Reidemeister et al, 2009; Marshall et al., 2010), with the currently favoured scenario featuring a mutually mean-motion resonant planetary system (Goździewski & Migaszewski, 2014).

In addition to the four giant planets, we now know that HR 8799 is orbited by at least two distinct debris belts – one interior to the orbit of HR 8799 e, and one exterior to the orbit of HR 8799 b (e.g. Rhee et al., 2007; Su et al., 2009; Matthews et al., 2014)[4]. The outermost of these is sufficiently bright and extended that it has been resolved in images taken by the Herschel Space Observatory (e.g. Matthews et al., 2014) and the Atacama Large Millimeter/submillimeter Array, ALMA (Booth et al., 2016). In contrast, the inner disc remains unresolved. As such its radial extent is poorly constrained, being based solely on the belt's thermal emission (e.g. Su et al. 2009, Reidemeister et al. 2009). In future, high contrast imaging may offer a potential avenue to resolve this component of the disc directly (using instruments such as SPHERE; Beuzit et al., 2008; Kasper et al., 2012), although this is complicated by the relatively face-on geometry of the system, which acts to reduce the observed surface brightness.

---

[4] HR 8799 is therefore one of a small number of debris disc systems that have spectral energy distributions best modelled by dust with two distinct temperatures. Kennedy & Wyatt (2014) studied a number of such systems, and examined whether they could be explained by the presence of a single narrow belt that produced the observed temperature range as a natural result of how the temperature of grains depends on their size. Whilst they found that single-ring models could theoretically fit the data, they also noted that the radially narrow annuli required are incompatible with observations for those cases where an observational determination of the disc structure has been possible. They therefore conclude that *'it is probable that most* [such discs] *have multiple spatial components'*.

In their detailed study of the HR 8799 system, in which they proposed the currently accepted multiply-resonant scenario for the orbits of the planets, Gozdziewski & Migaszewski (2014) also consider the dynamical stability of objects interior to the orbit of the innermost known planet (HR 8799 e). To do this, they simulated a swarm of a thousand test particles, distributed from the orbit of that planet inwards, to an inner edge of 2 au. This allowed them to produce a 'event time' graph for that region, showing that particles interior to about 8 au could survive on timescales of at least 160 Myr. Exterior to that distance, particles were highly dynamically unstable. Their work provided a first real glimpse of what the inner debris belt of HR 8799 might look like. They also considered the stability of Mars-mass bodies on million-year timescales through that region, resulting in a stability map (their Figure 22) that showed the key resonant structures that might sculpt the inner debris belt on those timescales.

In this work, we build upon the work of Gozdziewski & Migaszewski (2014), and attempt to determine the radial extent and likely structure of HR 8799's inner disc using detailed *n*-body dynamical simulations of the orbital evolution of two populations of 500,000 test particles, under the gravitational influence of the four known giant planets. In section two, we present the known parameters of the HR 8799 system, and restate the currently favoured model solution for the orbits of the known planets. In section three, we describe our simulations, before presenting our results in section four. We discuss our results and their implications in section five, before drawing together our conclusions in section six.

## 2. The HR 8799 system

HR 8799 is a young A-type star, thought to be a member of the Columba Association (e.g. Zuckerman et al., 2011). The precise age of HR 8799 remains poorly constrained. Zuckerman et al. (2013) suggest that the star is 30 million years (Myr) old, based on its likely membership of the Columba Association. In contrast, Reidemeister et al. (2009) suggest that it could be as old as 50 Myr. Whilst such a range of ages would not normally be problematic for studies of exoplanetary systems, in the case of HR 8799, the age plays a critical role. The older the system, the more the giant planets therein would have cooled since their formation, and therefore the more massive they would have to be in order to shine as brightly as required to explain the discovery observations detailed in Marois et al. 2008 and 2010.

For this reason, Marois et al. (2010) consider two different ages for the star when estimating the mass of its planet: 30 Myr, following a similar logic to Zuckerman et al., 2011, and 60 Myr, to maintain consistency with their earlier work on the system (Marois et al. 2008). In Table 1, we summarise the current best estimates of the stellar parameters for HR 8799, and where a discrepancy occurs, highlight those adopted in this work with an *.

| Parameter | Value | Reference |
| --- | --- | --- |
| Age (Myr) | 30<br>60* | Zuckerman et al., 2011<br>Marois et al., 2010 |
| Teff (K) | 7193 ± 87 | Baines et al., 2012 |
| L ($L_\odot$) | 5.05 ± 0.29 | Baines et al., 2012 |
| D (pc) | 39.40 ± 1.09 | Baines et al., 2012; following van Leeuwen, 2007 |
| M ($M_\odot$) | $1.516^{+0.038}_{-0.024}$<br>1.56* [5] | Baines et al., 2012<br>Goździewski & Migaszewski, 2014 |
| R ($R_\odot$) | 1.440 ± 0.06 | Baines et al., 2012 |
| [Fe/H] | -0.47 | Gray & Kaye, 1999 |
| log(g) | +4.23<br>+4.43 | Philip & Egret, 1980<br>David & Hillenbrand, 2015 |
| $P_{rot}$ (days) | 0.5053<br>0.3997 | Henry et al., 2007<br>Dubath et al., 2011 |

*Table 1. Stellar Parameters for HR 8799. Where multiple values are given, those marked by a * are the ones adopted for the simulations performed in this work.*

The HR 8799 system is located approximately 39 pc from the Earth (Van Leeuwen, 2007), and it is comprised of at least four giant planets and two debris belts. A recent dynamical study (Goździewski & Migaszewski, 2014) suggests that the planets are most likely trapped in a double Laplace resonance, with orbital period ratios of 1:2:4:8, after a rapid migration that followed their formation. They show that such a scenario results in the orbits of the planets being dynamically stable on long timescales, and, as such, argue that this is the most likely orbital architecture for the planetary system. In this work, we follow the results of Goździewski & Migaszewski (2014), with the best-fit orbital elements and estimated planetary masses detailed in Table 2.

| | $m$ [$m_{jup}$] | $a$ [au] | e | $i$ [deg] | $\Omega$ [deg] | $\omega$ [deg] | $M$ [deg] |
|---|---|---|---|---|---|---|---|
| HR 8799 e | 9±2 | 15.4±0.2 | 0.13±0.03 | 25±3 | 64±3 | 46±3 | 326±5 |
| HR 8799 d | 9±2 | 25.4±0.3 | 0.12±0.02 | | | 91±3 | 58±3 |
| HR 8799 c | 9±2 | 39.4±0.3 | 0.05±0.02 | | | 151±6 | 148±6 |
| HR 8799 b | 7±2 | 69.1±0.2 | 0.020±0.003 | | | 95±10 | 321±10 |

*Table 2. The orbital elements of HR 8799b, c, d and e, taken from Goździewski & Migaszewski (2014), as used in our integrations. Here, m denotes the mass of the planet, in Jupiter masses, a is the orbital semi-major axis, in au, and e is the eccentricity of the orbit. i is the inclination of the orbit to the plane of the sky, and $\Omega$, $\omega$ and $M$ are the longitude of the ascending node, the argument of pericentre, and the mean anomaly at the epoch 1998.83, respectively.*

In addition to the four planets, observations of the HR 8799 system at infrared wavelengths have revealed the presence of two distinct debris belts around the star (e.g. Rhee et al., 2007; Su et al., 2009; Matthews et al., 2014). These belts bracket the orbits of the planets, with one interior to the orbit of HR 8799 e (characterised as a 'hot' infrared excess), and the other exterior to the orbit of HR 8799 b. In addition to these belts, the entire system is swaddled by a halo of small grains of dust that stretch out to beyond 1000 au (Matthews et al., 2014). The outer debris belt, closely paralleling our cool Edgeworth-Kuiper belt, has been spatially resolved, and it is thought to extend from 100 (±10 au) to 310 au (Matthews et al., 2014). By contrast, the inner, warm belt, remains unresolved and poorly understood.

---

[5] Whilst the Baines et al. value for the mass of HR 8799 is the most precise determination available, we chose to use the value given in Gozdziewski & Migaszewski (2014) in our simulations for internal consistency, since the planetary masses and orbits we used were those presented in that work.

Based on a black-body fit to the observed temperature of the dust, the inner disc must lie within the orbit of HR 8799 e, with an outer edge at 15 au or less. Based on their analysis of the infrared excess around HR 8799, Reidemeister et al. (2009) note that truncating the outer edge of the inner belt at 10 au provides a slightly improved fit to the observational data than having the ring extend out to 15 au (the orbit of HR 8799 e), but find that the data are compatible with material stretching in all the way to at least 2 au. In contrast, Su et al. (2009) suggest that there is little evidence for dust interior to ~ 6 au in the system, this being the approximate location of the 4:1 mean-motion resonance with HR 8799 e. The idea of an outer edge to the disc at ~10 au is supported by Marois et al. (2010), who suggest that it can lie no more distant than that, based on the potential influence of HR 8799 e. Su et al. (2013) note that the hot dust is compatible with a radius of approximately 10 au, and estimate the dust production rate in the system assuming a belt centred at ~ 8 au.

With all this information taken in concert, it is clear that if HR 8799's outer belt is viewed as an analogue of our Edgeworth-Kuiper belt, then it is probably fair to consider the inner one to be similar to the Asteroid belt.

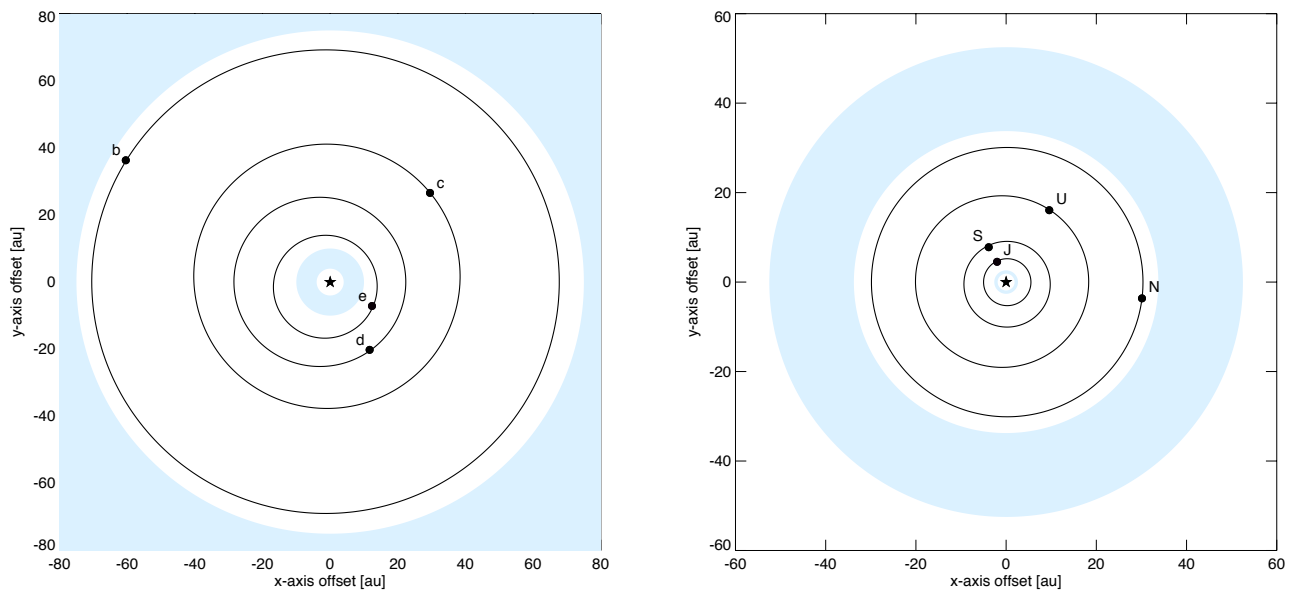

*Figure 2. Schematic plot of HR 8799 (left) in which the four planets are at their correct separations from the star, and locations around the star. The light blue regions correspond to the observed infrared excess described in Matthews et al. (2014). For comparison, we present the structure of our own Solar system, to the right, with the giant planets placed at their current orbital locations. The light blue regions here correspond to the Asteroid belt (interior to Jupiter, J), and the Edgeworth-Kuiper belt (exterior to Neptune, N).*

When all of this information is taken together, we can put together an illustrative schematic of the HR 8799 system, with its two debris belts and four giant planets, as presented in Figure 2. Note that, in that figure, the debris discs are shown in the locations inferred by the infrared observations described above. However, over the course of the evolution of the HR 8799 planetary system, the two belts will doubtless have been sculpted by the gravitational influence of the giant planets. In this work, we use that fact to constrain the location and structure of the inner, unresolved debris disc around HR 8799 through dynamical means, as described in the following section.

# 3. The Simulations

Given that the inner debris disc orbiting HR 8799 remains observationally unresolved, it is interesting to consider whether dynamical methods might be able to disentangle its structure, and place additional constraints on its radial extent. Such dynamical studies have previously been used to study the sculpting of the Solar system's small body populations by the migration of the giant planets (e.g. Lykawka et al., 2009, 2010, 2011; Lykawka & Horner 2010). In recent years, such studies have also allowed us to obtain better constraints on the orbits of recently discovered exoplanets than could be achieved on the basis of observation alone (e.g. Wittenmyer, Horner & Tinney, 2012; Wittenmyer et al., 2014c).

In this work, we therefore performed the most detailed simulations to date of the inner debris belt of HR 8799 using the Hybrid integrator within the n-body dynamics package MERCURY (Chambers, 1999), using UNSW's *Katana* and iVec's *Epic* supercomputing facilities, and building on our earlier work studying the system (Marshall, Horner & Carter, 2010; Contro et al., 2015a, b). We examined the orbital evolution of test particles that started our simulations at semi-major axes between 1 and 10 au (with an estimated outer boundary following the thermal modelling of the observed disc described in Su et al., 2009, and Matthews et al., 2014, and the estimated outer edge given in Marois et al., 2010). Given the wide range of orbital eccentricities and inclinations exhibited by the Solar system's stable small bodies, we attempted to constrain the maximum possible excitation for the HR 8799 disc. We therefore ran two discrete integrations, each following the evolution of a disc initially composed of 500,000 massless test particles under the influence of the four giant planets orbiting HR 8799 for a period of 60 Myr (to match the age assumed for the planetary system in Marois et al., 2008 and 2010). Our first integration followed the evolution of a disc of particles that was initially dynamically cold, whilst the second followed an excited, dynamically hot disc.

For each test particle in our two populations, each of the six orbital elements was randomly allocated within a set range. For both populations, the semi-major axes were randomly chosen to lie between 1 and 10 au, and each of the rotational orbital elements ($\omega$, $\Omega$ and M) was independently randomly assigned a value between 0 and 360 degrees. For our dynamically cold disc, particle eccentricities were distributed randomly between 0 and 0.1, and inclinations between 0 and 5 degrees. For the dynamically hot disc, the eccentricities ranged from 0.1 to 1.0, and the inclinations from 0 to 25 degrees, with an additional requirement that the initial pericentre distance for a test particle should be no smaller than 0.1 au[6]. The end results of these processes were a dynamically excited and a dynamically cold disc of debris that filled the inner reaches of the HR 8799 system.

The test particles created in this way were placed in planetary systems that included the four giant planets currently known to orbit HR 8799, using the parameters presented in Goździewski & Migaszewski, 2014), as detailed in Table 2. Each planet was initially placed on its nominal best-fit orbit, as detailed in that table, with the best-fit mass, and their orbits then evolved in our simulations under their mutual gravitational interaction. Our simulations adopted a stellar mass of $1.56 M_\odot$ for HR 8799, as shown in Table 1.

The dynamical evolution of the test particles was followed for a period of 60 Myr, with an integration time-step of 7 days, under the influence of the four giant planets. The 60 Myr run-time was chosen to ensure that our simulations were compatible with the longer of the estimated ages for the system (Marois et al., 2010). Each test particle was followed until it was either ejected from the planetary system (upon reaching a barycentric distance of 1000 au), or collided with one of the giant planets or the central star. The orbital elements of all surviving test particles were recorded every 6 Myr so that the temporal evolution of the disc's extent and architecture could be examined.

---

[6] This dynamically hot disc was included in our simulations in order to map the maximum boundaries of the HR 8799 debris belt in eccentricity and inclination space, and also allowed us to examine the evolution of objects in the system that would be comparable to the Solar system's near-Earth asteroid and short-period comet populations.

# 4. Simulation Results

In Figure 3, we present the temporal evolution of the dynamically cold disc, showing the instantaneous semi-major axes and eccentricities of all the test particles at the start of our simulations, and after 6, 30, and 60 Myr. Figure 4 shows the temporal evolution of the dynamically hot disc, again at the same intervals. Each disc initially comprised 500,000 test particles, although the number surviving decayed with time as the giant planets stirred the discs. Both figures reveal that the initial clearing of the most unstable regions of the disc was very rapid, with the discs taking on a form representative of their final structure within the first 6 Myr. After this, the sculpting of the discs slows, but continues throughout the period of integration, leading to a broadening of the gaps in the discs, and the further clearing of unstable test particles.

In Figure 5, we present the final distribution of surviving test particles for the two discs considered in this work, as a function of semi-major axis and eccentricity. The locations of several mean-motion resonances with HR 8799 e are plotted as vertical dashed red lines. Here, too, we plot the approximate region of the contemporaneous classical "Habitable Zone" (hereafter HZ; the region around the star in which an Earth like planet might reasonably be expected to be able to host liquid water on its surface, e.g. Kasting, Whitmire & Reynolds, 1993) with a vertical green bar. We estimated the current approximate boundaries of the HZ to be 1.974 (inner edge) to 3.407 au (outer edge), following Selsis et al. (2007) and Kopparapu et al. (2013), using the stellar parameters shown in Table 1. Figure 6 shows the final distributions of the surviving test particles as a function of semi-major axis and inclination. Although the sculpting in inclination space is less striking than that in eccentricity, it is clear that particles are dynamically stirred throughout the outer half of the debris belt, in some cases to extremely high inclinations.

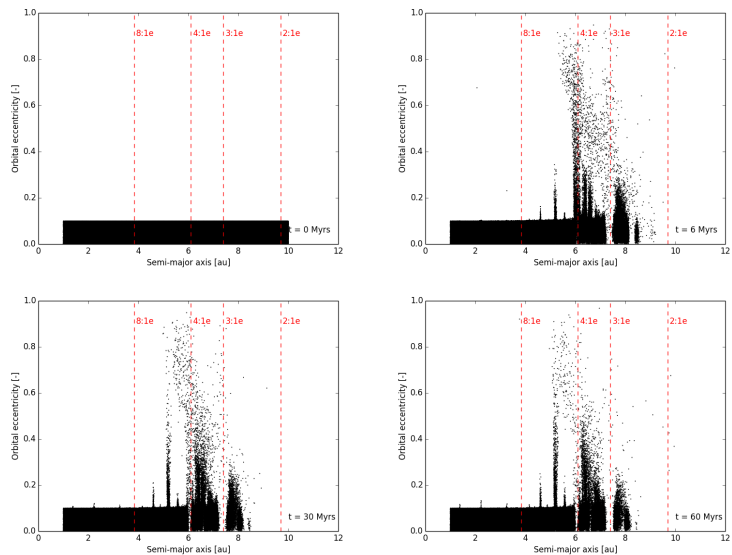

***Figure 3.*** *The evolution of a dynamically cold disc of debris orbiting HR 8799, under the influence of the four giant planets currently known in that system, for a period of 60 Myr. The panels show the initial conditions (top left), and then the surviving debris after 6 Myr (top right), 30 Myr (lower left), and finally 60 Myr (lower right), plotted in semi-major axis-eccentricity space. The red vertical dashed lines show the locations of the centres of various mean-motion resonances associated with the innermost of HR 8799's known planets, HR 8799 e. Note the rapid initial dispersal of debris beyond ~8.5 au, followed by a more sedentary sculpting of the disc over the remaining duration of the integration.*

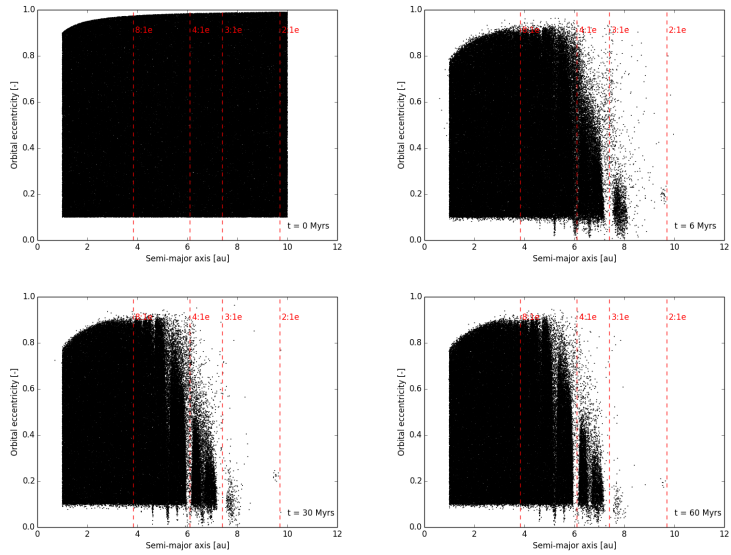

***Figure 4.*** *The evolution of a dynamically hot disc of debris orbiting HR 8799, under the influence of the four giant planets currently known in that system, for a period of 60 Myr. The panels show the initial conditions considered (top right), and then the surviving debris after 6, 30 and 60 Myr, plotted in semi-major axis-eccentricity space. The red vertical dashed lines show the locations of the centres of various mean-motion resonances associated with the innermost of HR 8799's known planets, HR 8799 e. As was the case for the dynamically cold disc (Figure 3), there is an initial rapid dispersal of debris beyond ~8 au, followed by a more gradual ongoing sculpting of the disc.*

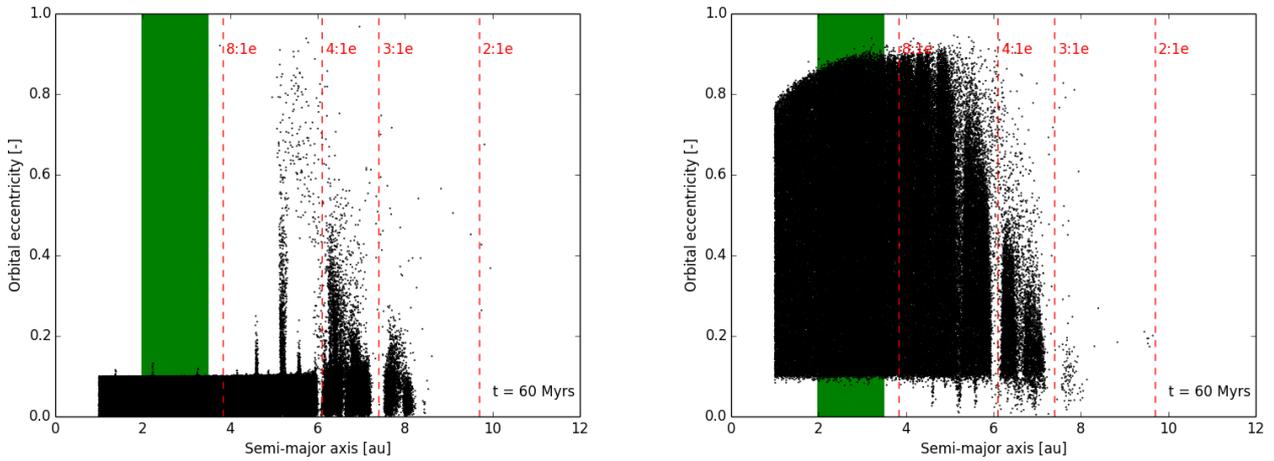

***Figure 5.*** *The final distribution of test particles orbiting HR 8799 in semi-major axis – eccentricity space after 60 Myr for dynamically cold (left) and hot (right) discs. The sculpting influence of mean-motion resonances between debris and planets can be clearly seen both in the gaps introduced to both discs (e.g. just outside 6 au), and in those locations where test particles have been driven to stable orbits with lower eccentricities that they initially possessed (the downward "spikes" visible throughout the lower plot).*

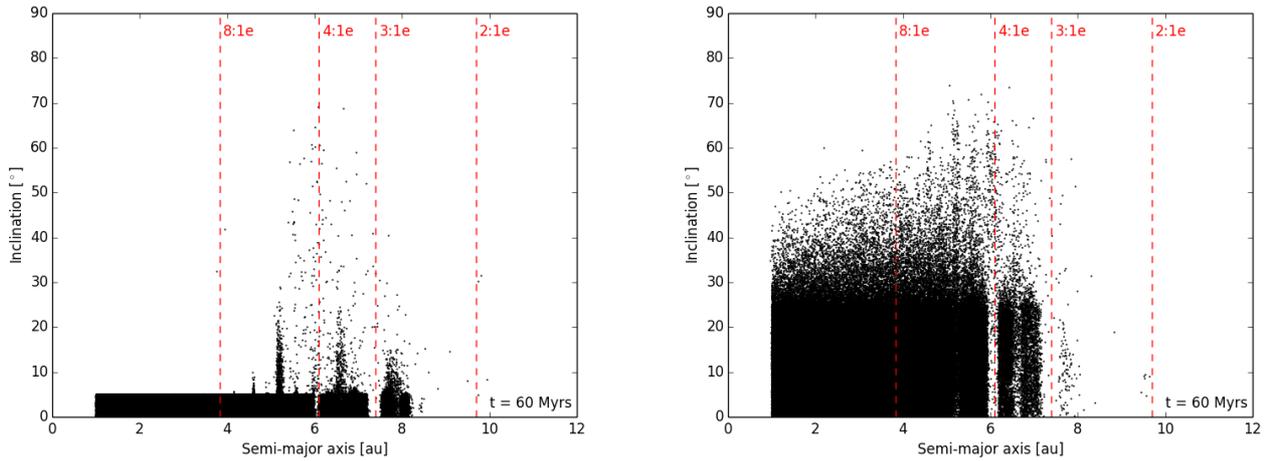

*Figure 6. The final distribution of test particles orbiting HR 8799 in semi-major axis – inclination space after 60 Myr for dynamically cold (left) and hot (right) discs. The strong sculpting influence of mean motion resonances can be seen throughout the outer half of the debris disc, with particles excited to large inclinations at barycentric distances greater than ~5 au, for the dynamically cold disc, and throughout the dynamically hot disc.*

## 5. Discussion of *n*-body Simulations

As can be readily seen in Figures 4 and 5, the outer edge of HR 8799's inner debris disc is strongly truncated by the gravitational influence of HR 8799 e. Within just 6 Myr of the start of our integrations, almost all test particles located exterior to ~8 au had been removed for both dynamically cold and dynamically hot initial discs, and they underwent a rapid initial sculpting. Debris in these regions are dynamically unstable on very short timescales, being rapidly removed from the system – just as objects that stray beyond the outer edge of the asteroid belt (orbital semi-major axis ~3.5 au) are quickly removed from our own Solar system as a result of the gravitational influence of Jupiter, unless they are trapped in one of the planet's mean motion resonances.

By the end of the simulations, the clearing of the region beyond ~8 au in the HR 8799 system is essentially complete. This result places a strong new dynamical constraint on the maximum radial extent of HR 8799's inner debris disc of just ~8 au.

The dynamical influence of HR 8799 e is also rapidly visible at smaller barycentric distances, with broad gaps being cleared in the disc at the location of that planet's 3:1 and 4:1 mean-motion resonances (and, to a lesser extent, around the 7:2 mean-motion resonance). Even between these resonances, the strong influence of the giant planets in the HR 8799 system can be clearly seen, exciting the objects in the dynamically cold disc to relatively high eccentricity and inclination. Indeed, the region between the 4:1 mean-motion resonance (at ~6 au) and the disc's outer edge (~8 au) strongly resembles our own Asteroid belt – Kirkwood gaps, excited objects and all.

The degree to which objects on initially 'dynamically cold' orbits are excited in this region is particularly interesting, since it naturally gives a mechanism by which the observed dust in the inner system is produced. If you have a dynamically cold population of objects, then the collision rate and mean collision velocity will be relatively small. However, if that same population becomes excited (both in inclination and eccentricity), then the mean collision velocity and collision rates will climb, as the disc collisionally grinds itself apart. As such, this region of excited but 'stable' objects is a good fit to the observed infrared excess.

Interior to the 4:1 mean-motion resonance, the stirring and clearing of the disc is much less pronounced – although additional gaps can be seen cleared in the dynamically hot disc, at high eccentricities, that correspond to the 5:1, 6:1 and 7:1 mean-motion resonances. These features are accompanied by downward 'spikes' in the distribution of test particles in that population, and 'upward' spikes in the dynamically cold distribution, showing that even distant resonant perturbations can induce significant excursions in orbital eccentricity and inclination. Interestingly, this mirrors a result found by Lykawka & Mukai (2007), who found that objects being transported outwards from the trans-Neptunian belt through perturbations at perihelion were sometimes trapped into distant, high-order mean-motion resonances with Neptune, following which they random walked down to low eccentricity orbits as a result of the resonant perturbations. For those trans-Neptunian objects, this 'resonant sticking' process allowed objects that escaped from the trans-Neptunian belt to survive for far longer than would otherwise be the case. Here, we are instead observing the opposite phenomenon – objects becoming dynamically excited by the influence of a distant, high-order resonance, until eventually many escape from the disc entirely.

Upon examination of the final distribution of dynamically hot objects (Figure 5), it is clear that, outside of the influence of mean-motion resonances, the outer edge of the disc is a strong function of orbital eccentricity. The more eccentric the orbit, the smaller an object's semi-major axis must be to survive for the full duration of our integrations. This is not a surprise – indeed, the outer edge of the disc roughly follows a line of constant apocentre. Objects above or to the right of that line come sufficiently close to the orbit of HR 8799 e at apastron that they experience strong chaotic perturbations, and eventually close encounters with that planet. Below or starward of the line, objects are sufficiently distant from HR 8799 e at apastron that they can remain dynamically stable on long timescales.

The results shown in the lower panel of Figure 5 also reveal where the validity of our simulations began to break down. At the inner edge of the dynamically hot disc, a clear sculpting can be seen at high eccentricities. This feature has nothing to do with the distant influence of the planets orbiting HR 8799. Instead, that feature is the result of objects moving on orbits with such small periastra that the 7-day time-step used for these integrations is insufficient. In other words, although striking, that feature is solely a computational effect, rather than the result of a real dynamical process.

Aside from the narrow bands influenced by mean-motion resonances, the dynamically cold disc interior to HR 8799 e's 4:1 mean-motion resonance remains unstirred. This fits nicely with the observations discussed in Su et al., 2009, who suggest that little dust is present interior to ~6 au, on the basis of their fit to the observed spectral energy distribution for the inner disc. This opens up an interesting possibility. Interior to 6 au, the giant planets are sufficiently distant as to not stir the disc. In other words, planetesimals in this region would not experience elevated collision velocities, meaning that the formation of the giant planets would not necessarily inhibit the formation of telluric worlds closer in.

Test particles within the HZ of HR 8799 remain almost entirely un-excited over the course of the simulations, a hint that planets may have been able to form in the inner HR 8799 system, and remain to be detected in the future. This does not, of course, infer that HR 8799 would be a good place to target for the search for life (e.g. Horner & Jones, 2010)! Even if an 'Earth-like' planet was found within the HZ, the system is likely far too young for any detectable life to have evolved on such a planet. Indeed, the very bright infrared excess seen around HR 8799 suggests that the debris belts therein are several orders of magnitude more massive than those in our Solar system – which would in turn suggest that any interior planets would be subject to an impact regime far more intense than that experienced by the Earth. Whilst this might help to deliver volatiles to any planets that formed in that inner region (e.g. Owen & Bar-Nun, 1995; Morbidelli et al., 2000; Horner et al., 2009), it would most likely render them uninhabitable at the current epoch. Indeed, such a high impact rate might even be considered to be the tail end of the accretion of objects in the inner reaches, suggesting that any telluric worlds in construction therein may not yet be complete – a result in keeping with the timescale considered for the formation of the terrestrial planets in our Solar system (e.g. Kenyon & Bromley, 2006). It is worth noting that the oldest minerals found to date on the Earth, zircons found in the Australian Jack Hills, date back to ~4.37 – 4.40 Gyr ago (e.g. Wilde et al., 2001; Valley et al., 2014), some 150 – 200 Myr after the Solar system's birth. With this representing the earliest confirmed crustal material on Earth, it may even be the case that any nascent telluric planets in the HR 8799 system will still be molten – and therefore certainly not habitable!

# 6. Dynamical MEGNO maps

In addition to our *n*-body simulations, we performed a complementary study of the structure of the inner debris belt based on the Mean Exponential Growth factor of Nearby Orbits (MEGNO, e.g. Cincotta & Simó, 2000; Goździewski et al., 2001; Cincotta, Giordano, C. M. & Simó, C., 2003) technique, allowing us to quantitatively measure the degree of chaos in the dynamical time evolution of the system. This technique is well tested, and has found widespread application within the field of dynamical astronomy (e.g. Hinse et al., 2010, Hinse et al., 2014). Considering a range of parameter space in semi-major axis and eccentricity of the test-masses we map the stability of our region of interest.

For the MEGNO analysis, we considered the restricted 6-body problem with a test-mass representing a member of the debris belt's planetesimal population. Thus the gravitational influence of HR8799 and its four confirmed planets on the test-mass were considered. The initial conditions of the massless particles are identical to those used for the full *n*-body simulations described in section 3.

To differentiate between chaotic and quasi-periodic time evolution of an object's orbit, the equations of motion and associated variational equations of motion are solved in parallel in order to calculate the variational vector and its time derivative at each integration time-step. We used the ODEX[7] time-variable multi-step algorithm as an independent method to solve the corresponding first-order equations of motion.

To quantify the degree of chaoticity in the motion of a test-mass as derived above, we calculate the MEGNO factor, $<Y>$. For an osculating initial condition resulting in a quasi-periodic time evolution, $<Y>$ asymptotically approaches 2. For aperiodic (chaotic) dynamics $<Y>$ diverges exponentially away from 2. The rate of divergence usually depends on the degree of perturbation inherent to the system. For practical reasons, we terminate a given integration as soon as $<Y> > 12$. For quasi-periodic motion, we usually find that $|<Y> - 2.0| < 0.0005$ at the end of the integration time.

The maps themselves were computed by varying either the initial semi-major axis, eccentricity or inclination of the test mass. This produced (a,e) and (a,i) MEGNO maps. For the (a,e) maps, the initial orbit of the test mass taken as being coplanar with the system's planets. For the (a,i) maps, we set the initial eccentricity to three different values (0.0, 0.05 and 0.25). For each test particle considered in this mapping process, the initial nodal longitude was set to zero, and the initial argument of pericentre and initial mean anomaly were set to 266 and 124 degrees, respectively – values chosen randomly at the start of the mapping process. In order to ensure independence from our earlier results, and to most efficiently use our available resources, we carried out the numerical MEGNO computations on a different computing cluster to that used for the *n*-body simulations described earlier. Our results can be seen in Figures 8 to 10, below.

In most cases, quasi-periodic motion is usually identified as an indicator of stable or regular motion, and chaotic motion is associated with unstable systems, usually resulting in collisions or ejections. However, inferring the qualitative character of dynamical systems has limitations. As is the case for every numerical indicator of chaos, a quasi-periodic orbit is only determined to be quasi-periodic *up to the integration time*. The dynamical character of the system for longer times is not known, and could in principle eventually turn out to be chaotic and possibly unstable.

---

[7] http://www.unige.ch/~hairer/prog/nonstiff/odex.f

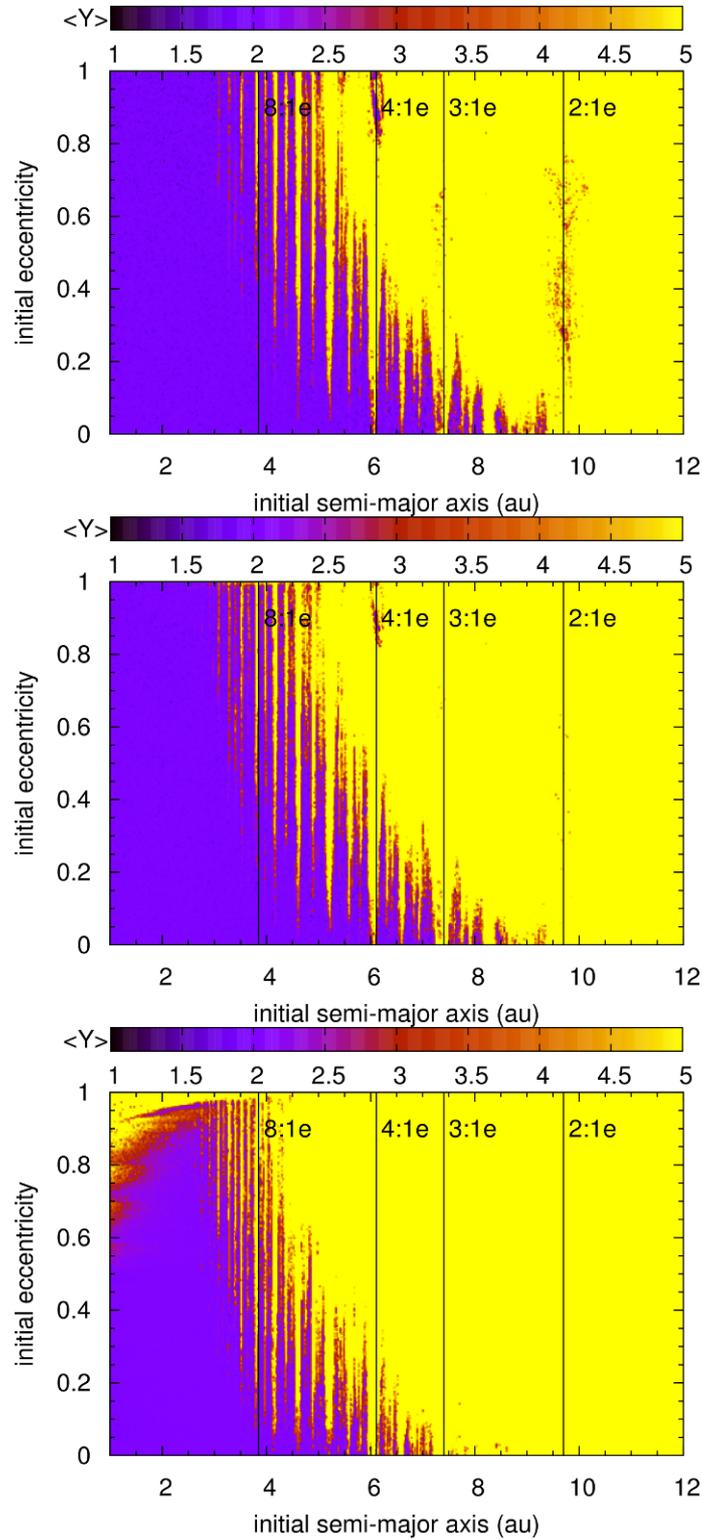

*Figure 7: Dynamical MEGNO maps for 30,000 test particles interior to the orbit of HR 8799 e integrated over 45 kyr, 100 kyr and 1 Myr (top to bottom). The orbit of each test particle was considered to be coplanar with the system, defining the longitude of the ascending node to be zero. The initial argument of pericentre and the mean anomaly were set to 266 and 124 degrees, respectively. Regions shown in yellow indicate chaotic or aperiodic orbital dynamics, whilst those in purple colours indicate quasi-periodic or regular dynamics. Vertical structures are mainly due to chaotic mean-motion resonances between the particles and the inner planet. The region interior to ~8 au is characterised by quasi-periodic stable motion for an initially cold disc.*

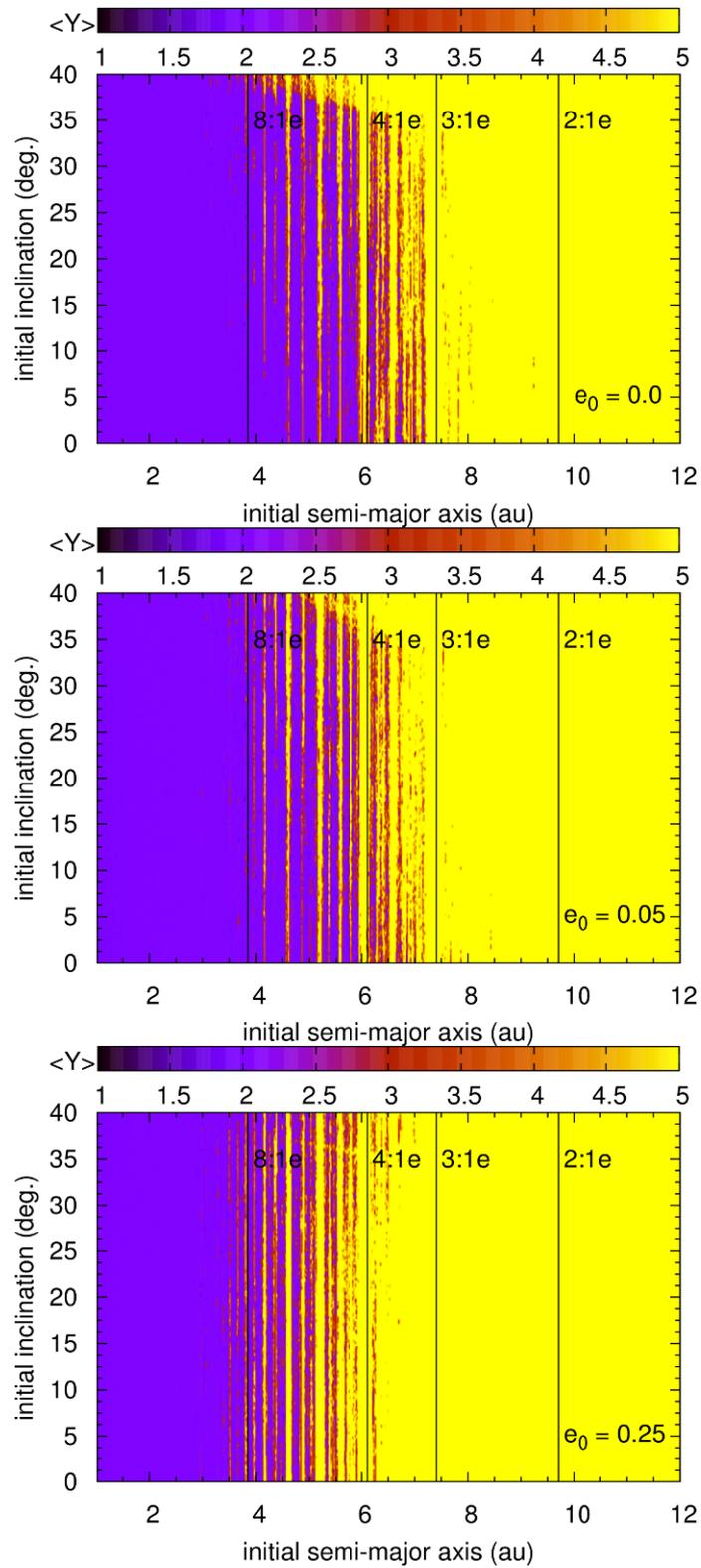

*Figure 8: MEGNO maps of the chaoticity of the region interior to the orbit of HR 8799 e, as a function of the initial orbital eccentricity of the test particles considered. The upper panel shows the results for particles on initially circular orbits, whilst the middle and lower panels show the results for eccentricities of 0.05 and 0.25, respectively. In each case, the simulations here spanned a period of 1 Myr.*

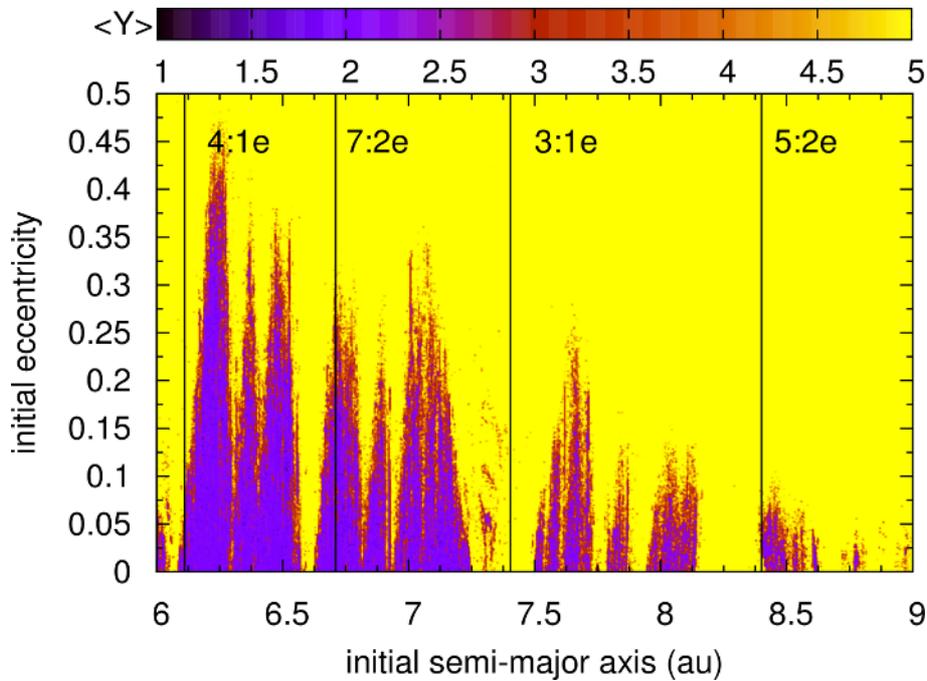

*Figure 9: A close-up of the structure evidence in the results of a 100 kyr MEGNO mapping of the region likely occupied by the inner HR8799 debris belt. Here, the fine structure induced in the belt by the influence of mean motion resonances can be clearly seen, revealing a structure that is in many ways analogous that observed in our own Asteroid belt.*

## 7. MEGNO discussion

We show the results of our dynamical MEGNO mapping process in Figures 8 and 9, displaying the chaoticity of the orbits of test particles within the HR 8799 system as a function of their initial orbital elements. In general, we considered semi-major axes between 1 and 12 au, eccentricities between 0 and 1, and inclinations between 0 and 40 degrees. Our calculations were performed with a [400:600] grid in the orbital elements of interest. Our results are colour coded such that regions of chaotic behaviour ($<Y> > 2$) are shown in yellow, whilst quasi-periodic behaviours are shown in purple.

In Figure 7, we show the results for three different integration times – namely 45 kyr, 100 kyr and 1 Myr. Comparing our results to those shown in Figure 6, we find qualitatively good agreement between the two different methods. In particular, we observe that MEGNO reproduces the 6:1 and 4:1 mean-motion resonance between debris belt particles and the innermost planet, which yield chaotic evolution on even the shortest timescales considered.

Higher-order mean-motion resonances can be seen to become important for low-eccentricities over longer timescales, and are even more influential as the initial inclination excitation of the test particles is increased. As the timescale of the integrations continues, the chaoticity of the outer regions of the disc increases markedly. This highlights the different dynamical timescales on which instability occurs, and reflects the temporal evolution that can clearly be seen in Figure 3 and 5. Over longer timescales still, it is likely that the debris belt would continue to be whittled away, as those regions that evolve on longer dynamical timescales are gradually depleted of material – a process that will no doubt be exacerbated by the on-going collisional grinding of the objects therein that is producing the observed infra-red excess.

In the lower panel of Figure 7, we can also see the long-term effect of eccentricity pumping for particles that initial start on highly-eccentric orbits (upper-left corner). As particle orbits in this region evolve, and their eccentricities are pumped, the pericentre distance of the test particles shrinks. As a result, many are eventually removed via collision with the central body. Such behaviour is observed in the evolution of small bodies within our Solar system, with the Kreutz sungrazing comets being a prime example of the continuing accretion of small bodies by the Sun (e.g. Marsden, 1967, 2005).

In Figure 8, we present MEGNO maps covering the semi-major axis vs. orbital inclination space for three different initial test particle eccentricities, chosen to span the range between dynamically cold and hot debris belts. Once again, the strong influence of mean-motion resonances on the evolution of the debris belt can be clearly seen at all chosen orbital inclinations. Once again, our results are in good agreement with the direct multi-particle simulations, with the outer edge of the inner debris belt being found to be lie at around 8 au.

Finally, we would like to highlight a subtle detail in the dynamical topology displayed in Figure 8. When one compares our results to those displayed in Figure 22 (top panel) of Goździewski & Migaszewski (2014), it is clear that we do not find a stable periodic island within the 2:1e mean-motion resonance. The reason for this difference is most likely because we calculate MEGNO maps for fixed initial values of the orbital rotation angles. By contrast, Goździewski & Migaszewski (2014) consider a wide variety of combinations for these angles. If at least one set of orbital parameters renders the system to be stable at a given grid point, then the authors assign stable dynamics to that particular location. This demonstrates that system stability is often a highly complex issue, with quasi-periodic orbits that exist only for certain particular combinations of initial orbital parameters – the investigation of which would be an interesting topic for future study for this system.

A zoomed map of the region between semi-major axes of 6 and 9 au and eccentricities between 0 and 0.5 is shown in Figure 9. This plot reveals a significant amount of fine structure that, at least superficially, resembles the structure of our own Asteroid belt. Small islands of quasi-periodic motion can be clearly seen in the regions between those destabilised by the influence of mean-motion resonances. These resonances are the root-cause of the striking dynamical sculpting of the debris belt right on the border of the outer truncation region of the disc.

## 8. Simulating the collisional environment in the inner HR 8799 system

To build on our earlier suite of dynamical integrations, and the results of our MEGNO mapping procedure, we carried out one final suite of simulations. These runs were designed to examine the collisional environment that would exist in the inner reaches of the HR 8799 system and the debris belt itself.

To do this, we took the full list of test particles that survived for the full 60 Myr of our integration time, and filtered those survivors to extract two populations of object – those that would make up the debris belt itself, and those which would correspond to analogues of the Solar system's short period comets and near-Earth asteroids.

To achieve this, we made a simple cut on our dataset of surviving particles. We kept all those particles which had pericentre or apocentre between 6 au and 10 au. In addition, we kept any particles that had an orbital eccentricity of 0.2 or higher. In total, this gave us a population of 396,682 test particles, taken from both the hot and cold disc simulations detailed above.

We then integrated the evolution of those test particles forward in time under the gravitational influence of the system's four planets for a period of 100 years, recording their instantaneous location and velocities 1000 times through the course of the simulation. This resulted in just under four hundred million particle locations and velocities – essentially producing spatial density and velocity grid for our two populations of object.

We then calculated the collision density resulting from this swarm of data, under the assumption that any planetary objects interior to the inner edge of the debris belt would move on orbits coplanar with the giant planets. Since these hypothetical 'interior planets' were not themselves modelled, we instead checked for all occurrences within our dataset when test particles lay within 0.1 au of the plane of the planetary orbits, and considered these potential impacts. For those impacts, we calculated the collision velocities that would occur upon an object travelling on a circular orbit – neglecting, of course, the acceleration that would occur due to the mass of the objects colliding.

This process allowed us to move from our final distribution of test particles to a 'collision probability profile' for the inner system, build on the implicit assumption of a debris belt, analogous to the asteroid belt, with an inner edge at 6 au, and populations of unstable objects sourced from that region, analogous to our Solar system's near-Earth asteroids and short-period comets.

In the Solar system, the great majority of the near-Earth asteroids are dynamically de-coupled from strong perturbations by the giant planets – as is comet 2P/Encke (e.g. Steel & Asher, 1996, Horner et al., 2003, Levison et al., 2006). These objects are thought to have entered their current orbits through the combination of gravitational perturbations (including the influence of the $v_6$ secular resonance, at the inner edge of the Asteroid belt, e.g. Morbidelli et al., 1994, Froeschle et al., 1995, Migliorini et al., 1997) and non-gravitational effects (including collisions, and the Yarkovsky-YORP effect, e.g. Morbidelli & Vokrouhlický, 2003, O'Brien, D. P. & Greenberg, R., 2005, Bottke et al., 2005, 2006).

Since our simulations included no non-gravitational forces, and therefore no route by which such objects could be created, it was important to include particles from the dynamically hot disc simulations to provide the near-Earth asteroid analogues for this collisional work. This inherently introduces something of a bias, since there are many such objects in our runs that, in the absence of collisions and non-gravitational effects, will survive in the inner HR 8799 system for the full duration of our simulations. As such, this will naturally lead to the impact/collision rate in the inner regions of the HR 8799 being over-estimated, to some degree, when compared to the collision rate in the debris belt. Despite this, our results can provide new insights to the collisional regime that might be expected in the inner reaches of the HR 8799 system.

## 9. Collisional Results and Discussion

The results of our collisional mapping simulations are presented in Figures 10 and 11. Figure 10 presents the results for the entire inner system, at distances between 1 and 10 au, whilst Figure 11 zooms in on the debris belt itself, the region between 6 and 8 au that might well be analogous to the asteroid belt within our own Solar system.

It is immediately apparent that the highest collision rate is to be found within the debris belt itself. This is, of course, not particularly surprising, since that was the region with by far the highest particle density. Equally apparent is that the great majority of the collisions in that region will be relatively slow – with collisions between objects in the belt having typical velocities of just 1.2 kms$^{-1}$, or less. It is interesting to note that this low collision velocity is actually less than the mean collision velocity expected between asteroids in the Solar system's asteroid belt (e.g. <V> = 5.81 ± 1.88 kms$^{-1}$; Farinella & Davis, 1992; <V> = 5.3 kms$^{-1}$; Bottke et al., 1994). This is a natural result of the larger scale of the HR 8799 when compared to the Solar system – at a distance roughly three times greater from the central body than the asteroids in our Solar system, HR 8799's asteroids will move more slowly, and so collisions would be expected to typically be somewhat more gentle.

Despite this, such collisions will certainly be sufficient to disrupt asteroid-sized bodies in this region. For example, recent hydrocode simulations collisions between large objects (Movshovitz et al., 2016) show that even 100 km radius basaltic and icy asteroids could be catastrophically disrupted in collision with 50 km projectiles with collision velocities of significantly less than 1 kms$^{-1}$. They find that disruption is expected when the kinetic energy involved in a collision exceeds ~2 to 8 times the gravitational binding energy of the target + impactor system. Our results therefore suggest that the debris belt as modelled in our earlier simulations would be undergoing a process of collisional attrition, rather than accretion, a finding wholly consistent with the observation of a warm infrared excess from the system's inner reaches.

Interior to the debris belt, the potential collision rate is significantly lower, despite the fact that the region is overpopulated in our collisional runs when compared to the debris belt itself, due to the lack of non-gravitational and collisional means of particle removal in our initial runs. Within the Solar system, the number of near-Earth asteroids larger than a kilometre in diameter is estimated to be ~990 (Harris & D'Abramo, 2015). The main belt asteroids are more numerous than this by at least a factor of a thousand (e.g. Tedesco & Desert, 2002). Were we to assume that the two populations scale comparably in the HR 8799 system, and use this to extrapolate from our results, the collision rates in the inner region would be several orders of magnitude lower than in the debris belt.

Collisions in this region, driven by particles on eccentric orbits, can be significantly more energetic, although at all distances, most collisions occur with velocities less than ~15 kms$^{-1}$. The paucity of truly high-velocity collisions (like those between Leonid meteoroids and the Earth, at ~71 kms$^{-1}$, e.g. Cook, 1973; Beech, 1998; McBeath & Arlt, 2000) is the result of the test particles moving on prograde orbits. None of the 396,682 test particles that survived our initial integrations had evolved onto retrograde orbits – which in turn means that the collisions modelled by our 100 year simulations were all in a prograde direction.

These collisions would be analogous to those experienced by the terrestrial planets from the near-Earth asteroids – which themselves originate in the main asteroid belt (e.g. Gladman, Michel & Froeschlé, 2000; Morbidelli & Vokrouhlický, 2003), and are the product of collisions between the asteroids. Should planets exist interior to the HR 8799 inner debris belt, then, it is likely that they will receive an ongoing impact flux from the collisional fragments produced by grinding in the debris belt – just as the terrestrial planets continue to experience in our own Solar system.

The margin by which collisions are more frequent in the debris belt itself than interior to its inner edge would in turn suggest that, barring cometary activity from massive Centaur-like comets in the inner HR 8799 system, the great bulk of dust produced therein would originate within the debris belt itself – which is consistent with the observations and modelling of the dust emission from the inner, hot belt (e.g. Su et al. 2009, Reidemeister et al. 2009).

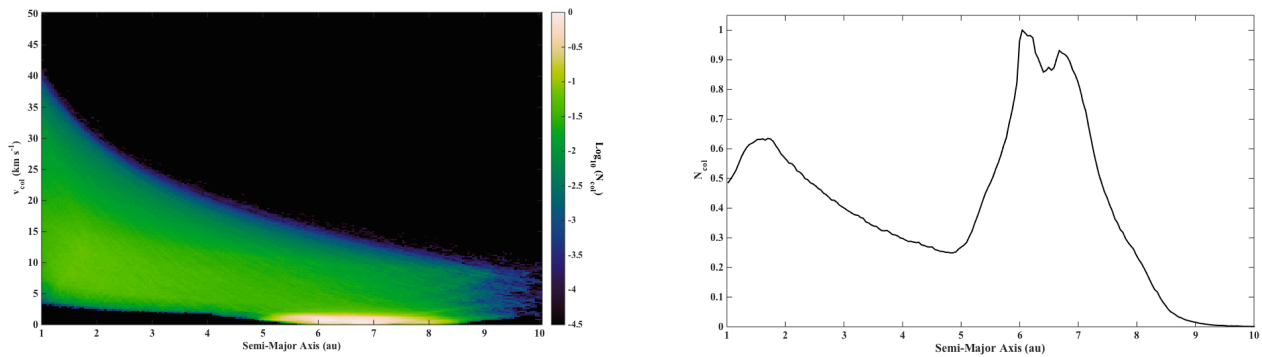

*Figure 10: The distribution of collisions that would result in the HR 8799 system from the test particles that survived the full 60 Myr of our simulations within the main belt of the system (between ~6 and ~8.5 au), and from the comet and near-Earth asteroid analogues that remained in the inner system at that point. The left panel shows the log of the normalised number of collisions as a function of semi-major axis and collision velocity across the region studied. The right panel shows simply the normalised number of collisions as a function of location in that region.*

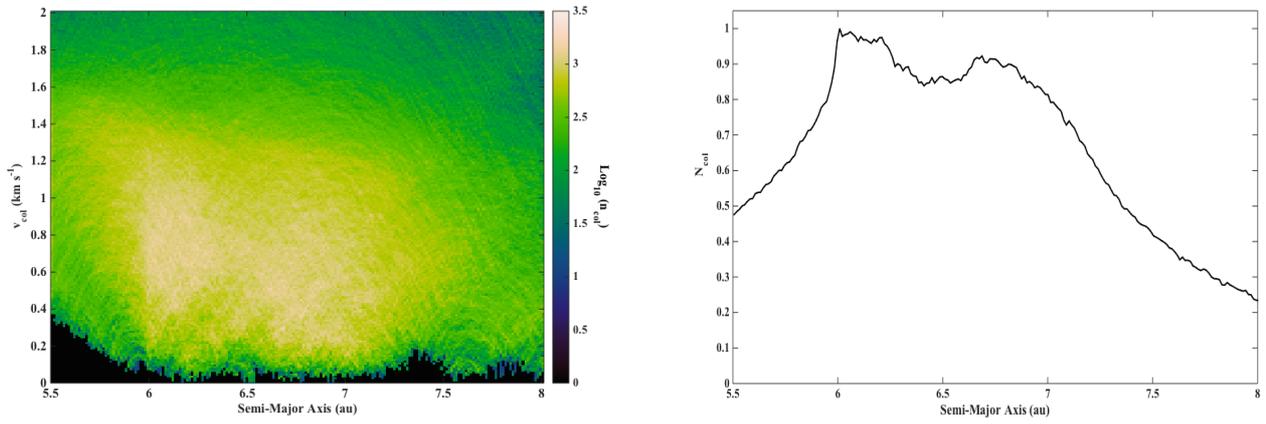

*Figure 11: The collision rate in the main part of the HR 8799 'asteroid belt' modelled in our earlier simulations. The left panel is a zoomed in version on region that appears brightest in Figure 10, showing collisions between members of the belt itself, whilst the right panel shows the total normalised collision rate as a function of location in the disc.*

## 10. Conclusions

We have carried out highly detailed dynamical simulations of the inner debris disc in the HR 8799 planetary system, building on earlier work by Goździewski & Migaszewski (2014). By examining the evolution of one million test particles over a period of sixty million years, we find that the debris belt must be sharply truncated at a semi-major axis of ~8 au, as predicted in that earlier work. Our results were complemented by the production of MEGNO stability maps, which support our conclusion that the outer edge of the belt can lie no further than ~8 au from HR 8799.

Interior to this, our simulations support the conclusions of Goździewski & Migaszewski (2014), revealing that the debris belt is likely highly structured, just like our own Asteroid belt. The debris belt exhibits features reminiscent of the Kirkwood gaps, and regions of stability where members are nonetheless excited to moderate eccentricities and inclinations, even if they formed on dynamically cold orbits. That excitement, in turn, increases the collision rate and mean collision velocity between objects in the disc, providing a natural explanation as to why the infrared excess observed is best modelled as originating in that region. Simply put – that is where we would expect the production of dust to be greatest.

This conclusion is supported by subsidiary simulations, using the test particles that survived the full 60 Myr duration of our initial simulations. Those simulations reveal that the debris belt would be collisionally active, experiencing an ongoing collisional attrition much like our own asteroid belt. In contrast to the asteroid belt, where the mean collision velocity is somewhat in excess of 5 $kms^{-1}$ (e.g. Farinella & Davis, 1992; Bottke et al., 1994), we find that the majority of asteroid-asteroid collisions in the HR 8799 inner debris belt would occur with velocities of order 1.2 $kms^{-1}$ or less. Despite this, such collisions remain in the destructive regime, and would explain the observed warm infrared excess observed in the system.

Interior to HR 8799's 4:1 mean-motion resonance, the disc is markedly less stirred (aside from narrow bands corresponding to more distant, high order resonances). This, in turn, suggests that this region is sufficiently dynamically stable, and sufficiently distant from the massive planets, for the formation of telluric worlds to not be ruled out.

As such, our findings support those of Su et al., 2009, who suggest that the inner edge of HR 8799's 'asteroid belt' likely lies at the location of the 4:1 mean-motion resonance. It may well be that, in the future, nascent terrestrial planet analogues are found in the HR 8799 system, interior to its asteroid belt.

Such a discovery would further enhance the HR8799 system's claims to be a true Solar system analogue – albeit younger, bigger, and far more violent.

## Acknowledgements

This work is supported by School of Physics - UNSW Australia, and made use of UNSW Australia's *Katana* and iVec's *Epic* supercomputing facilities. This research has made use of the Exoplanet Orbit Database and the Exoplanet Data Explorer at exoplanets.org, and NASA's Astrophysics Data System (ADS). JH gratefully acknowledges the support of USQ's Strategic Research Fund: the STARWINDS project. JPM is supported by a UNSW Vice Chancellor's postdoctoral research fellowship. Our numerical computations also made use of the SFI/HEA Irish Center for High-End Computing (ICHEC) and the POLARIS computing cluster at the Korea Astronomy and Space Science Institute (KASI). TCH acknowledges support through KASI research grant #2015-1-850-04. The authors wish to thank the referee, Cezary Migaszewski, for their helpful and insightful feedback, which led to significant improvements in our work.

## References

Alvarez, L. W., Alvarez, W., Asaro, F., & Michel, H. V., 1980, Science, 208, 1095

Aumann, H. H., Beichman, C. A., Gillett, F. C., et al., 1984, ApJ, 278, L23

Aumann, H. H., 1985, PASP, 97, 885

Backman, D. E. & Paresce, F., 1993, Protostars and Planets III, 1253

Baines, E. K., White, R. J., Huber, D. et al., 2012, ApJ, 761, 57

Batalha, N. M., Rowe, J. F., Bryson, S. T. et al., 2013, ApJS, 204, 24

Becker, L., Poreda, R. J., Hunt, A. G., Brunch, T. E., & Rampino, M., 2001, Science, 291, 1530

Beech, M., 1998, The Astronomical Journal, 116, 499

Beichman, C. A., Bryden, G., Stapelfeldt, K. R. et al., 2006, ApJ, 652, 1674

Beuzit, J.-F., Feldt, M., Dohlen, K. et al., 2008, Ground-based and Airborne Instrumentation for Astronomy II. Edited by McLean, Ian S.; Casali, Mark M. Proceedings of the SPIE, Volume 7014, article id. 701418

Boisse, I., Pepe, F., Perrier, C. et al., 2012, A&A, 545, 55

Booth, M., Kennedy, G., Sibthorpe, B. et al., 2013, MNRAS, 428, 1263

Booth, M., Jordán, A., Casassus, S. et al., 2016, MNRAS, 460, L10

Borucki, W. J., Koch, D., Basri, G. et al., 2010, Science, 327, 977

Borucki, W. J., Koch, D. G., Basri, G. et al., 2011, ApJ, 736, 19

Bottke, W. F., Nolan, M. C., Greenberg, R. & Kolvoord, R. A., 1994, Icraus, 107, 255

Bottke, W. F., Durda, D. D., Nesvorný, D., Jedicke, R., Morbidelli, A., Vokrouhlický, D. & Levison, H. F., 2005, Icarus, 179, 63

Bottke, W. F., Vokrouhlický, D., Rubincam, D. P. & Nesvorný, D., 2006, Annual Review of Earth and Planetary Sciences, 34, 157

Brown, M. E., 2001, AJ, 121, 2804

Burns, J. A., Lamy, P. L. & Soter, S., 1979, Icarus, 40, 1

Chambers, J. E., 1999, MNRAS, 304, 793

Cincotta, P. M. & Simó, C., 2000, Astronomy & Astrophysics Supplement, 147, 205

Cincotta, P. M., Giordano, C. M. & Simó, C., 2003, Physica D, 182, 151

Contro, B., Wittenmyer, R. A., Horner, J. & Marshall, J. P., 2015, Origins of Life and Evolution of Biospheres, 45, 41

Contro, B., Wittenmyer, R. A., Horner, J. & Marshall, J. P., 2015, Proceedings of the 14$^{th}$ Australian Space Research Conference, University of South Australia, Adelaide, 29$^{th}$ September – 1$^{st}$ October 2014; ISBN: 13: 978-0-9775740-8-7; Editors Wayne Short & Iver Cairns

Cook, A. F., 1973, NASA Special Publication, 319, 183

David, T. J. & Hillenbrand, L. A., 2015, ApJ, 804, 146

Delfosse, X., Bonfils, X., Forveille, T. et al., 2013, Astronomy & Astrophysics, 553, 15

Dones, L., Weissman, P. R., Levison, H. F., & Duncan, M. J., 2004, Comets II, M. C. Festou, H. U. Keller, and H. A. Weaver (eds.), University of Arizona Press, Tucson, 745, 153

Dubath, P., Rimoldini, L., Süveges, M. et al., 2011, MNRAS, 414, 2602

Duchêne, G., Arriaga, P., Wyatt, M. et al., 2014, ApJ, 784, 148

Eiroa, C., Marshall, J. P., Mora, A., et al., 2013, Astronomy & Astrophysics, 555, 11

Ertel, S., Marshall, J. P., Augereau, J.-C. et al., 2014, Astronomy & Astrophysics, 561, 114

Faramaz, V., Beust, H., Thébault, P. et al., 2014, Astronomy & Astrophysics, 563, 72

Farinella, P. & Davis, D. R., 1992, Icarus, 97, 111

Froeschle, C., Hahn, G., Gonczi, R., Morbidelli, A. & Farinella, P., 1995, Icarus, 117, 1, 45

Fukagawa, M., Itoh, Y., Tamura, M. et al., 2009, ApJ, 696, L1

Gladman, B., Kavelaars, J. J., Petit, J.-M., Morbidelli, A., Holman, M. J. & Loredo, T., 2001, AJ, 122, 1051

Gladman B., Michel, P. & Froeschlé, C., 2000, Icarus, 146, 176

Goździewski, K., Bois, E., Maciejewski, A. J. & Kiseleva-Eggleton, L., 2001, Astronomy & Astrophysics, 378, 569

Goździewski, K. & Migaszewski, C., 2014, MNRAS, 440, 3140

Gray, R. O. & Kaye, A. B., 1999, AJ, 118, 2993

Habing, H. J.; Dominik, C.; Jourdain de Muizon, M.; et al., 2001, Astronomy and Astrophysics, 365, 545

Han, E., Wang, S. X., Wright, J. T., Feng, Y. K., Zhao, M., Fakhouri, O., Brown, J. I & Hancock, C., 2014, Publications of the Astronomical Society of the Pacific, 126, 827

Harris, A. W. & D'Abramo, G., 2015, Icarus, 257, 302

Henry, G. W., Fekel, F. C. & Henry, S. M., 2007, The Astronomical Journal, 133, 1421

Hinse, T. C., Christou, A. A., Alvarellos, J. L. A. & Goździewski, K., 2010, MNRAS, 404, 837

Hinse, T. C., Horner, J. & Wittenmyer, R. A., 2014, JASS, 31, 187

Horner, J., Evans, N. W., Bailey, M. E. & Asher, D. J., 2003, MNRAS, 343, 1057

Horner, J., & Jones, B. W., 2008, International Journal of Astrobiology, 7, 251

Horner, J., & Jones, B. W., 2009, International Journal of Astrobiology, 8, 75

Horner, J., Mousis, O., Petit, J. –M., & Jones, B. W., 2009, Planetary and Space Science, 57, 1338

Horner, J. & Jones, B. W., 2010, International Journal of Astrobiology, 9, 273

Howard, A. W., Johnson, J. A., Marcy, G. W. et al., 2010, ApJ, 721, 1467

Howard, A. W., Marcy, G. W., Bryson, S. T. et al., 2012, ApJS, 201, 20

Jewitt, D., Luu, J., 1993, Nature, 362, 730

Jewitt, D. C., Trujillo, C. A. & Luu, J. X., 2000, AJ, 120, 1140

Jones, H. R. A., Butler, R. P., Tinney, C. G. et al., 2010, MNRAS, 403, 1703

Kasper, M., Beuzit, J.-L., Feldt, M. et al., 2012, The Messenger, 149, 17

Kasting, J. F., Whitmire, D. P. & Reynolds, R. T., 1993, Icarus, 101, 108

Kennedy, G. M. & Wyatt, M. C., 2013, MNRAS, 433, 2334

Kennedy, G. M. & Wyatt, M. C., 2014, MNRAS, 444, 3164

Kenyon, S. J. & Bromley, B. C., 2006, The Astronomical Journal, 131, 1837

Kopparapu, R. K., Ramirez, R. M., Schottel-Kote, J. et al., 2013, ApJ, 765, 131

Krivov, A. V., 2010, Research in Astronomy and Astrophysics, 10, 5, 383

Krivov, A. V., Reidemeister, M., Fiedler, S., Löhne, T, Neuhäuser, R., 2011, MNRAS, 418, L15

Lagrange, A.–M., Vidal-Madjar, A., Deluil, M., Emerich, C., Beust, H. & Ferlet, R., 1995, Astromy & Astrophysics 296, 499

Lagrange, A.–M., Gratadour, D., Chauvin, G. et al., 2009, Astronomy and Astrophysics, 493, L21

Lestrade, J. -F., Matthews, B. C., Sibthorpe, B., et al., 2012, Astronomy & Astrophysics, 548, 86

Levison, H. F., Terrell, D., Wiegert, P. A., Dones, L. & Duncan, M. J., 2006, Icarus, 182, 161

Liseau, R., Eiroa, C., Fedele, D. et al., 2010, Astronomy & Astrophysics, 518, 132

Löhne, T, Augereau, J. –C., Ertel, S. et al., 2012, Astronomy & Astrophysics, 537, 110

Lykawka, P. S. & Mukai, T., 2007, Icarus, 192, 238

Lykawka, P. S., Horner, J., Jones, B. W., & Mukai, T., 2009, MNRAS, 398, 1715

Lykawka, P. S., Horner, J., Jones, B. W., & Mukai, T., 2010, MNRAS, 404, 1272

Lykawka, P. S. & Horner, J., 2010, MNRAS, 405, 1375

Lykawka, P. S., Horner, J., Jones, B. W., & Mukai, T., 2011, MNRAS, 412, 537

Macintosh, B., Graham, J. R., Barman, T. et al., 2015, Science, 350, 64

Marois, C., Macintosh, B., Barman, T. et al., 2008, Science, 322, 1348

Marois, C., Zuckerman, B., Konopacky, Q. M. et al., 2010, Nature, 468, 1

Marsden, B. G., 1967, Astronomical Journal, 72, 1170

Marsden, B. G., 2005, ARA&A, 43, 75

Marshall, J., Horner, J. & Carter, A., 2010, International Journal of Astrobiology, 9, 259

Marshall, J. P., Löhne, T, Montesinos, B. et al., 2011, Astronomy & Astrophysics, 529, 117

Marshall, J. P., Kirchschlager, F., Ertel, S., et al., 2014, Astronomy & Astrophysics, 570, 114

Matthews, B. C., Sibthorpe, B., Kennedy, G. et al., 2010, Astronomy & Astrophysics, 518, 135

Matthews, B. C., Kennedy, G., Sibthorpe, B. et al., 2014, ApJ, 780, 12

Mayor, M. and Queloz, D., 1995, Nature, 378, 355-359

McBeath, A. & Arlt, R, 2000, WGN, Journal of the International Meteor Organization, 28, 4, 91

Migliorini, F., Morbidelli, A., Zappala, V., Gladman, B. J., Bailey, M. E. & Cellino, A., 1997, Meteoritics and Planetary Science, 32, 7, 903

Minton, D. A., & Malhotra, R., 2009, Nature, 457, 1109

Morbidelli, A., Gonczi, R., Froeschle, C. & Farinella, P., 1994, Astronomy and Astrophysics, 282, 955

Morbidelli, A., Chambers, J., Lunine, J. I. et al., 2000, Meteoritics and Planetary Science, 35, 1309

Morbidelli, A. & Vokrouhlický, D., 2003, Icarus, 163, 120

Morbidelli, A., Brasser, R., Gomes, R., Levison, H. F. & Tsiganis, K., 2010, AJ, 140, 1391

Movshovitz, N., Nimmo, F., Korycansky, D. G., Asphaug, E. & Owen, J. M., 2016, Icarus, 275, 85

Mullally, F., Coughlin, J. L., Thompson, S. E. et al., 2015, ApJSS, 217, 31

Nesvorný, D., & Ferrz-Mello, S., 1997, Icarus, 130, 247

Neugebauer, G., Habing, H. J., van Duinen, R. et al., 1984, ApJ, 278, 1

O'Brien, D. P. & Greenberg, R., 2005, Icarus, 178, 179

Owen, T. & Bar-Nun, A., 1995, Icarus, 116, 215

Patel, R. I., Metchev, S. A. & Heinze, A., 2014, ApJSS, 212, 10

Pawallek, N., Krivov, A. V., Marshall, J. P. et al., 2014, ApJ, 792, 65

Philip, A. G. D. & Egret, D., 1980, Astronomy and Astrophysics Supplement Series, 40, 199

Pilbratt, G. L., Riedinger, J. R., Passvogel, T. et al., 2010, Astronomy and Astrophysics, 518, 1

Rauer, H., Catala, C., Aerts, C. et al., 2014, Experimental Astronomy, 38, 249

Reidemeister, R., Krivov, A. V., Schmidt, T. O. B. et al., 2009, Astronomy and Astrophysics, 503, 247

Ricker, G. R., Winn, J. N., Vanderspek, R. et al., 2015, Journal of Astronomical Telescopes, Instruments, and Systems, 1, 014003

Robertson, P., Horner, J., Wittenmyer, R. A. et al., 2012, ApJ, 754, 50

Rowe, J. F., Bryson, S. T., Marcy, G. W. et al., 2014, 784, 45

Selsis, F., Kasting, J. F., Levrard, B. et al., 2007, Astronomy & Astrophysics, 476, 1373

Sheppard, S. S. & Trujillo, C. A., 2006, Science, 313, 511

Stark, C. C., Roberge, A., Mandell, A. & Robinson, T. D., 2014, ApJ, 795, 122

Steel, D. I. & Asher, D. J., 1996, MNRAS, 281, 937

Swift, J. J., Bottom, M., Johnson, J. A. et al., 2015, Journal of Astronomical Telescopes, Instruments, and Systems, 1, 027002

Su, K. Y. L., Rieke, G. H., Stansberry, J. A. et al., 2006, ApJ, 653, 675

Su, K. Y. L., Rieke, G. H., Stapelfeldt, K. R., 2009, ApJ, 705, 314

Su, K. Y. L., Rieke, G. H., Malhotra, R. et al., 2013, ApJ, 763, 118

Takeda, G., Ford, E. B., Sills, A. et al., 2007, ApJS, 168,297

Tedesco, E. F., Desert, F.-X., 2002, AJ, 123, 2070

Trilling, D. E., Bryden, G., Beichman, C. A. et al., 2008, ApJ, 674, 1086

van Leeuwen, F., 2007, Astrophysics and Space Science Library, 350

Valley, J. W., Cavosie, A. J., Ushikubo, T. et al., 2014, Nature Geoscience, 7, 219

Wilde, S. A., Valley, J. W., Peck, W. H. & Graham, C. M., 2001, Nature, 409, 175

Wittenmyer, R. A., Horner, J., Tuomi, M. et al., 2012, ApJ, 753, 12

Wittenmyer, R. A., Horner, J. & Tinney, C. G., 2012, ApJ, 761, 165

Wittenmyer, R. A., Tinney, C. G., Horner, J. et al., 2013, PASP, 125, 351

Wittenmyer, R. A., Tuomi, M., Butler, R. P., et al., 2014a, ApJ, 791, 11

Wittenmyer, R. A., Horner, J., Tinney, C. G. et al., 2014b, ApJ, 783, 103

Wittenmyer, R. A., Tan, X., Lee, M. H. et al., 2014c, ApJ, 780, 140

Wyatt, S. P. & Whipple, F. L., 1950, The Astrophysical Journal, 111, 134

Wyatt, M. C., 2008, ARAA, 46, 339

Wyatt, M. C., Clarke, C. J. and Booth, M., 2011, Celestial Mechanics and Dynamical Astronomy, 111, 1-2, 1

Zechmeister, M., Kürster, M., Endl, M. et al., 2013, A&A, 552, 78

Zuckerman, B., Rhee, J. H., Song, I. & Bessell, M. S., 2011, ApJ, 732, 61